# The Optimal Lattice Quantizer in Nine Dimensions

*Bruce Allen\* and Erik Agrell*

The optimal lattice quantizer is the lattice that minimizes the (dimensionless) second moment $G$. In dimensions 1 to 3, it has been proven that the optimal lattice quantizer is one of the classical lattices, and there is good numerical evidence for this in dimensions 4 to 8. In contrast, in 9 dimensions, more than two decades ago, the same numerical studies found the smallest known value of $G$ for a non-classical lattice. The structure and properties of this conjectured optimal lattice quantizer depend upon a real parameter $a > 0$, whose value was only known approximately. Here, a full description of this one-parameter family of lattices and their Voronoi cells is given, and their (scalar and tensor) second moments are calculated analytically as a function of $a$. The value of $a$ which minimizes $G$ is an algebraic number, defined by the root of a 9th order polynomial, with $a \approx 0.573223794$. For this value of $a$, the covariance matrix (second moment tensor) is proportional to the identity, consistent with a theorem of Zamir and Feder for optimal quantizers. The structure of the Voronoi cell depends upon $a$, and undergoes phase transitions at $a^2 = 1/2$, 1, and 2, where its geometry changes abruptly. At each transition, the analytic formula for the second moment changes in a very simple way. The methods can be used for arbitrary one-parameter families of laminated lattices, and may thus provide a useful tool to identify optimal quantizers in other dimensions as well.

## 1. Introduction and Summary

Lattices are regular arrays of points in $\mathbb{R}^n$. They are obtained as arbitrary linear combinations of $n$ linearly independent basis vectors, with integer coefficients. The remarkable book by Conway and Sloane[1] provides a comprehensive review of lattices and their properties.

An important geometric structure of lattices are their Voronoi cells,[2] named after Georges Voronoï who established much of the underlying theory.[3–6] Voronoi cells are also called Wigner–Seitz cells,[7] Brillouin zones,[8] or Dirichlet cells.[9] The Voronoi cell of a lattice point $p$ is the set of points in $\mathbb{R}^n$ which are closer to $p$ than to any other lattice point, with the conventional Cartesian/Euclidean metric and norm. All of the Voronoi cells have identical shape, and may be obtained by translation of the cell about the origin $p = 0$. They cover space, intersecting only on their boundaries, and thus have an $n$-volume which is the same as that of the parallelotope defined by the $n$ basis vectors.

Because the distance inequalities which define them are saturated on planes, Voronoi cells are convex polytopes with flat faces. The faces of the cell about the origin lie on planes that are halfway in between the origin and a set of nearby lattice points. The intersections of these faces form lower-dimensional convex sub-faces, also bounded by planes, which in turn intersect to form lower-dimensional faces. At the bottom of this hierarchy are the 0-faces, or vertices. These vertices define the Voronoi cell, in the sense that their convex hull is the cell. A single Voronoi cell also defines the lattice, because one can use translated copies to cover the space, thus identifying the lattice points, which lie at the center of the cells.

Lattices arise in many diverse fields, including number theory,[10] geometry,[11] cryptography,[12] string theory,[13] coding and information theory,[14] and data analysis.[15] In the latter two contexts, lattices which minimize the average squared distance to the closest lattice point[16] provide the "most efficient" solutions for accurately discretizing information content, or in searching for unknown signals. This average squared distance may be written as $U/V$, where $V$ is the volume of the Voronoi cell Equation (3.2) and $U$ is the un-normalized second moment, which is the trace of Equation (3.6). Since $U$ scales as (length)$^{n+2}$, the dimensionless quantity to be minimized for an $n$-dimensional lattice is[16,17]

$$G = \frac{1}{n} \frac{U}{V^{1+2/n}}. \tag{1.1}$$

(The normalization factor $1/n$ ensures that $G = 1/12$ for a cubic lattice $\mathbb{Z}^n$ in any number of dimensions.) Since the geometric shape that minimizes $G$ for a given volume is a ball, the optimal quantizer lattice is likely to have a Voronoi cell which is close to this shape.

B. Allen
Max Planck Institute for Gravitational Physics
Callinstrasse 38, Hannover 30167, Germany
E-mail: bruce.allen@aei.mpg.de

E. Agrell
Department of Electrical Engineering
Chalmers University of Technology
Gothenburg SE-41296, Sweden

The ORCID identification number(s) for the author(s) of this article can be found under https://doi.org/10.1002/andp.202100259

In dimensions from 1 to 3, the lattices that minimize $G$ have been found analytically (see Chapter 2 in ref. [1] and ref. [18]). This is done by brute force: the quadratic form for the squared distance (in lattice coordinates) is written as a function of several unknowns, and the average is minimized. For example, in 3 dimensions[18] the solution is the body-centered cubic lattice (bcc = $A_3^* = D_3^*$.) No optimality proofs are available for quantizers in dimensions higher than 3, but well-known classical lattices are conjectured to be optimal in dimensions up to 8 (see pp. 12 and 61 in ref. [1])

The strongest evidence for this is numerical. More than two decades ago, Agrell and Eriksson for the first time used computerized optimization algorithms to search for optimal lattice quantizers in higher dimensions.[19] Effectively they evaluated $G$ via Monte Carlo integration, shifted the lattice points using stochastic gradient descent, and continued until a stable minimum was found. The key advance which enabled this was an efficient algorithm[20] to identify the closest point of an arbitrary given lattice to any given point in $\mathbb{R}^n$. The process was repeated many times in each dimension $n = 3, 4, \ldots, 10$, each time optimizing the lattice over all $n(n+1)/2 - 1$ degrees of freedom, without imposing any constraints on the structure of the lattice. Most trials converged to the same lattices, up to equivalence operations. The numerically optimized generator matrices are available online.[21] Based on this numerical evidence, it was conjectured in ref. [19] that the obtained lattices are the globally optimal quantizers.

In dimensions $n < 9$, the numerically optimized lattices that were obtained were the same classical lattices that had already been conjectured as optimal in pp. 12 and 61 of ref. [1]. But in dimension 9, Agrell and Eriksson identified a previously unknown lattice, which is not a classical lattice, though it is related to the $E_8$ root lattice and very similar to $\Lambda_9^*$, which is the dual of the laminated lattice $\Lambda_9$. In this paper, we call this new lattice AE$_9$, after its discoverers.

The AE$_9$ lattice[22] is defined by lattice points $Z B$, where $Z$ is the set of nine-dimensional row vectors whose components are integers. The $9 \times 9$ generator matrix is

$$B = \begin{bmatrix} 2 & 0 & 0 & 0 & 0 & 0 & 0 & 0 & 0 \\ 1 & 1 & 0 & 0 & 0 & 0 & 0 & 0 & 0 \\ 1 & 0 & 1 & 0 & 0 & 0 & 0 & 0 & 0 \\ 1 & 0 & 0 & 1 & 0 & 0 & 0 & 0 & 0 \\ 1 & 0 & 0 & 0 & 1 & 0 & 0 & 0 & 0 \\ 1 & 0 & 0 & 0 & 0 & 1 & 0 & 0 & 0 \\ 1 & 0 & 0 & 0 & 0 & 0 & 1 & 0 & 0 \\ 1 & 0 & 0 & 0 & 0 & 0 & 0 & 1 & 0 \\ \frac{1}{2} & \frac{1}{2} & \frac{1}{2} & \frac{1}{2} & \frac{1}{2} & \frac{1}{2} & \frac{1}{2} & \frac{1}{2} & a \end{bmatrix}. \quad (1.2)$$

The upper left corner of (1.2) is a generator of the root lattice $D_8$, which means that the lattice consists of copies of the $D_8$ lattice, shifted and stacked along the ninth dimension. The projection of the nine-dimensional lattice orthogonal to the ninth dimension is another root lattice, $E_8$. In ref. [19], the quantity $a$ in the lower right hand corner was found by numerical optimization, yielding $a \approx 0.573$. The exact value was unknown, and might have even been a rational number.

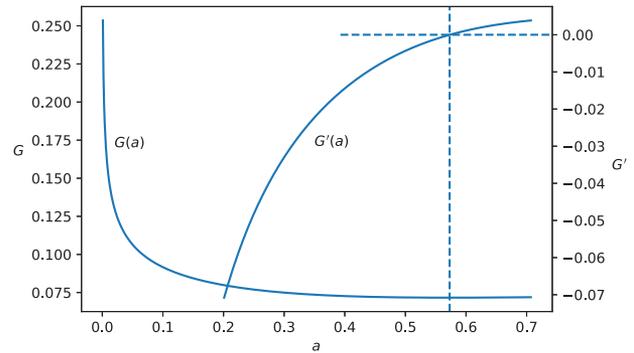

**Figure 1.** The second moment $G$ and its derivative.

In this paper, we calculate $G(a)$ of AE$_9$ analytically; a plot of this function is shown in **Figure 1**. The value of $a$ which minimizes $G$ is the algebraic number $a = \sqrt{v}$, where $v$ is the smallest positive root of the polynomial

$$720v^9 - 1704v^8 + 2544v^6 - 3528v^4 + 4896v^2 - 4320v + 929. \quad (1.3)$$

Thus $a = 0.5732237949\ldots$, corresponding to $G = 0.0716225944\ldots$. An interesting byproduct of this calculation is that we obtain a full analytic description of the Voronoi cells of AE$_9$.

In the course of this work we have explored the one-parameter family of lattices generated by $B$, obtaining exact formulae for the second moments, volumes, and other properties of all of the subfaces (face catalogs are included in the Supporting Information. There is one catalog for each of the four phases shown in Table 3 in Section 4). In nine dimensions, for $a^2 < 1/2$ the volume and second moment are:

$V = 2a$, and

$$U = -\frac{a^{19}}{90} + \frac{4a^{17}}{135} - \frac{8a^{13}}{135} + \frac{28a^9}{225} - \frac{16a^5}{45} + \frac{2a^3}{3} + \frac{929a}{810}. \quad (1.4)$$

This is obtained from an analytic and geometric description which is valid only for $a^2 < 1/2$. At the lower boundary of this interval, for $a = 0$, the nine-volume vanishes, and at the upper boundary the volumes of the faces $F_1^3$, $F_1^7$, $F_2^3$, $F_3^7$, $F_4^{10}$, and $F_5^{10}$ vanish. (We denote equivalence classes of congruent $n$-dimensional faces by $F_n^t$, where the "type label" $t$ identifies the class—see Table 2). However, provided that $a$ is within this interval, the overall topology of the Voronoi cell (by which we mean the number of faces in any dimension and their pattern of intersection) is unchanged, and our analysis and equations are valid.

As $a$ increases, the Voronoi cell undergoes three distinct phase transitions. While we do not present the other phases here, they are discussed in Section 4 and full details are provided in the face catalogs (see Supporting Information).

There is a beautiful theorem by Zamir and Feder, stating that a necessary (but not sufficient) condition for the optimal lattice quantizer is a covariance proportional to the identity matrix.[23] The covariance is proportional to the second moment tensor $U^{\mu\nu}$, defined in Equation (3.6) where $\mu$ and $\nu$ are vector indices that run from 1 to $n$. For our one-parameter family of lattices, symmetry implies[24] that

$$U^{\mu\nu} = \alpha(a)\delta^{\mu\nu} + \beta(a)z^\mu z^\nu, \qquad (1.5)$$

where the first term is proportional to the identity matrix $\delta^{\mu\nu} = \text{diag}(1,\ldots,1)$, and the second to the outer product of a pair of unit vectors along the 9th or "vertical" coordinate, $z^\mu = (0,\ldots,0,1)$. We find

$$\alpha = \frac{a^{19}}{90} - \frac{7a^{17}}{270} + \frac{a^{13}}{27} - \frac{7a^9}{150} + \frac{2a^5}{45} + \frac{929a}{6480}, \text{ and}$$

$$\beta = -\frac{a^{19}}{9} + \frac{71a^{17}}{270} - \frac{53a^{13}}{135} + \frac{49a^9}{90} - \frac{34a^5}{45} + \frac{2a^3}{3} - \frac{929a}{6480}. \qquad (1.6)$$

The trace of $U^{\mu\nu}$ is the scalar moment $U = 9\alpha + \beta$ of Equation (1.4). Multiplying by 6480 shows that $\beta z^\mu z^\nu$ vanishes for precisely the same value of $a$ that extremizes G. Were that not so, the Zamir and Feder result would imply that $AE_9$ is not the optimal lattice quantizer.

Currently, the best methods for computing Voronoi cells and their moments are those of Dutour Sikirić, Schürmann, and Vallentin,[25] with which they computed the quantizer constant for a number of lattices, including $\Lambda_9^*$. That lattice corresponds to setting $a = 1/2$, for which all entries along the bottom row of $B$ become $1/2$. Happily, our general formula reproduces their value (see Table 5 in ref. [25]), which is

$$G(\Lambda_9^*) = 1371514291/19110297600$$
$$= 0.0717683376\ldots. \qquad (1.7)$$

An interesting consequence of our construction is that it demonstrates that the structure of the $AE_9$ and $\Lambda_9^*$ Voronoi cells are the same[26] although some faces of the $\Lambda_9^*$ cell have additional congruences. Specifically, for $a = 1/2$, the one-faces $F_1^1$, $F_1^6$, and $F_1^7$ become identical. In higher dimensions there are a handful of faces which become congruent in shape, though with different overall scales: $F_2^1$, $F_2^3$, $F_2^6$ in dimension 2, $F_3^3$, $F_3^7$, $F_3^{11}$ in dimension 3, and $F_4^6$, $F_4^{10}$ in dimension 4.

As described below, our construction begins analytically, using "pencil and paper", and is completed using a small Python code that we wrote for this purpose. The code constructs the Voronoi cell faces in all dimensions, both in floating point, for specific values of $a$, and symbolically, for general values of $a$. As far as we know, other computer codes (for example Polyhedral, see refs. [25, 27]) are limited to rational arithmetic and therefore not directly applicable.

At the conclusion of the paper, in Section 6, we explain how it may be possible to use existing rational arithmetic codes and techniques to infer these results. More generally, our approach can be used to evaluate the scalar and tensor second moments $U$ and $U^{\mu\nu}$, the volume $V$ and the dimensionless scalar second moment $G$ for one-parameter families of lattices in any number of dimensions $n$. For this purpose, it should be sufficient to evaluate $U$ and $V$ for at most $2n + 3$ distinct rational values of the parameter.

## 2. Construction of the Voronoi Cell

The $AE_9$ Voronoi cell centered at the origin is the nine-face $F_9^1$. We begin by finding its facets (eight-faces) and vertices (zero-faces). For this, it is helpful to examine the symmetries implied by the form of the generator matrix $B$ of Equation (1.2).

The symmetries of the lattice which preserve the origin are a group of order 10 321 920, and most easily described in terms of three generator subgroups, which only overlap at the identity. (Our description is for generic values of $a$; if $a$ takes special values such as $1/2$, then there are additional symmetries and the group is larger.) $R$ (reflection): Suppose that $(x_1,\ldots,x_9)$ is a lattice point. Then $(x_1,\ldots,x_8,-x_9)$ must also be a lattice point. $R$ is a subgroup of order 2. $P$ (permutation): If $(x_1,\ldots,x_9)$ is a lattice point, then any permutation of the first eight coordinates is also a lattice point. $P$ is a subgroup of order $8! = 40\,320$. $S$ (sign inversion): If $(x_1,\ldots,x_9)$ is a lattice point, then changing the sign of any two of the first eight coordinates also produces a lattice point. $S$ is a subgroup of order 128. Since the lattice points have these symmetries, any geometric object defined by the lattice, and in particular the Voronoi cell about the origin, will share them.

To begin the construction of the Voronoi cell, we apply the algorithm in Section VI(C) of ref. [20] to identify the relevant vectors, that is, the lattice vectors that define the facets of the Voronoi cell. There are 370 such vectors, provided that $a^2 < 1/2$; the other cases are described in Section 4.

The 370 relevant vectors are of three types, with different distances to the origin. (For $a^2 \geq 2/15$, these relevant vectors correspond to the 370 closest lattice points to the origin, but not for $a^2 < 2/15$.) There are, respectively, 256, 112, and 2 lattice points of the three types; the facets lie halfway between the origin and these points. The coordinates of one point of each type are

$$n_1 = \frac{1}{2}(1,1,1,1,1,1,1,1,2a) \text{ with } |n_1|^2 = a^2 + 2,$$

$$n_2 = (1,1,0,0,0,0,0,0,0) \text{ with } |n_2|^2 = 2, \text{ and}$$

$$n_3 = (0,0,0,0,0,0,0,0,2a) \text{ with } |n_3|^2 = 4a^2. \qquad (2.1)$$

The symmetry operations listed above can be applied to these three lattice points to generate others at the same distance, and hence the other 367 facets.

The lattice vector $n_1$ is acted on by $R$ and $S$ to generate 256 points. The symmetry $R$ has no effect on $n_2$, $P$ provides $\binom{8}{2} = 28$ permutations, and $S$ provides 4, so $n_2$ generates 112 lattice points. For the third vector, neither $P$ nor $S$ have any effect, so $n_3$ provides just two lattice points under the action of $R$. The facets (eight-faces) lie in planes orthogonal to these 370 vectors, at distance $|n_i|/2$ from the origin.[28]

The vertices lie at the intersections of these facets and are listed in **Table 1**. To illustrate how they are found, we calculate the coordinates of one of these vertices; the others are obtained in similar fashion. Consider, for example, the vertex which lies at the intersection of the following nine facets: (1) the "diagonal" facet defined by $n_1$, (2–8) the seven "vertical" facets defined by permutations $P$ of $n_2$ which leave a "1" in the first coordinate, and (9) the "top" facet defined by $n_3$.

We start by intersecting the facets defined by $n_2$ and $n_3$. A point $(x_1,\ldots,x_9)$ that lies on both facets must be equidistant from the origin and the corresponding lattice points, and hence

**Table 1.** For $a^2 < 1/2$, the Voronoi cell has 93 024 vertices which fall into 16 equivalence classes under symmetry transformations. This table lists those classes $H_i$, ordered by decreasing distance $|H_i|$ from the origin, with the coordinates of a representative vertex. The ordering depends upon the parameter $a$, which we set to the special value $a \approx 0.573$ that minimizes the dimensionless second moment $G$. The last two columns are the number of vertices in the equivalence class, and the number of facets of each type ($F_8^1$, $F_8^2$, $F_8^3$) that meet at the vertex.

| Class | Vertex | | | | | | | | | $w$ | $|H_i|^2$ | Number | Facets |
|---|---|---|---|---|---|---|---|---|---|---|---|---|---|
| $H_1$ | ( 0, | 0, | 0, | 0, | 0, | 0, | 0, | 1, | $a$) | | $1 + a^2$ | 32 | (0,14,1) |
| $H_2$ | ($-w$, | $w$, | $w$, | $w$, | $w$, | $w$, | $w$, | $1-w$, | $a$) | $(1-a^2)/4$ | $1 + a^2/2 + a^4/2$ | 2048 | (7,7,1) |
| $H_3$ | ($-w$, | $w$, | $w$, | $w$, | $w$, | $w$, | $w$, | $1-w$, | 0) | $(1+a^2)/4$ | $1 + a^2/2 + a^4/2$ | 1024 | (14,7,0) |
| $H_4$ | ( $w$, | $w$, | $w$, | $w$, | $w$, | $w$, | $w$, | $1-w$, | $a$) | $(1-a^2)/6$ | $(8 + 8a^2 + 2a^4)/9$ | 2048 | (1,7,1) |
| $H_5$ | ($-w$, | $w$, | $w$, | $w$, | $w$, | $w$, | $w$, | $w$, | 0) | $(2+a^2)/6$ | $(8 + 8a^2 + 2a^4)/9$ | 128 | (16,0,0) |
| $H_6$ | ($-w$, | $w$, | $w$, | $w$, | $w$, | 1/2, | 1/2, | 1/2, | 0) | $(1+2a^2)/6$ | $(8 + 5a^2 + 5a^4)/9$ | 7168 | (10,3,0) |
| $H_7$ | ( 0, | 0, | 0, | $w$, | $w$, | $w$, | $w$, | $1-w$, | $a$) | $(1-a^2)/3$ | $(8 + 5a^2 + 5a^4)/9$ | 17 920 | (4,4,1) |
| $H_8$ | ( 0, | 0, | 0, | 0, | $w$, | $w$, | $w$, | $1-w$, | $a$) | $(1-a^2)/2$ | $1 + a^4$ | 8960 | (8,3,1) |
| $H_9$ | ( 0, | 0, | 0, | $w$, | 1/2, | 1/2, | 1/2, | 1/2, | 0) | $a^2$ | $1 + a^4$ | 8960 | (8,6,0) |
| $H_{10}$ | ( 0, | 0, | 0, | 0, | $w$, | 1/2, | 1/2, | 1/2, | $a$) | $(1-2a^2)/2$ | $1 + a^4$ | 8960 | (8,3,1) |
| $H_{11}$ | ($-w$, | $w$, | $w$, | $w$, | 1/2, | 1/2, | 1/2, | 1/2, | 0) | $a^2/2$ | $1 + a^4$ | 8960 | (8,6,0) |
| $H_{12}$ | ($-w$, | $w$, | $w$, | $w$, | $w$, | 1/2, | 1/2, | 1/2, | $a$) | $(1-2a^2)/6$ | $(8 + 4a^2 + 5a^4)/9$ | 14 336 | (5,3,1) |
| $H_{13}$ | ( 0, | 0, | 0, | 0, | $w$, | $w$, | $w$, | $1-w$, | 0) | $(1+a^2)/3$ | $(8 + 4a^2 + 5a^4)/9$ | 8960 | (8,4,0) |
| $H_{14}$ | ( 0, | 0, | 0, | 0, | 1/2, | 1/2, | 1/2, | 1/2, | $a/2$) | | $1 + a^2/4$ | 2240 | (8,6,0) |
| $H_{15}$ | ( $w$, | $w$, | $w$, | $w$, | $w$, | $w$, | $w$, | $1-w$, | 0) | $(1+a^2)/6$ | $(8 + a^2 + 2a^4)/9$ | 1024 | (2,7,0) |
| $H_{16}$ | ($-w$, | $w$, | $w$, | $w$, | $w$, | $w$, | $w$, | $w$, | $a$) | $(2-a^2)/6$ | $(8 + a^2 + 2a^4)/9$ | 256 | (8,0,1) |

must satisfy the equations

$$x_1^2 + \cdots + x_9^2 = x_1^2 + \cdots + x_8^2 + (x_9 - 2a)^2, \text{ and}$$
$$x_1^2 + \cdots + x_9^2 = (x_1 - 1)^2 + (x_2 - 1)^2 +$$
$$x_3^2 + \cdots + x_9^2. \qquad (2.2)$$

All the quadratic terms drop out, leaving two linear equations. The first implies that $x_9 = a$ and the second implies that $x_1 + x_2 = 1$. Now consider the intersection of this with additional facets of type $n_2$, but permuted by $P$ in a way that leaves a "1" in the first coordinate of $n_2$ but shifts the second "1" further along. These give equations of identical form to $x_1 + x_2 = 1$, but with $x_2$ replaced by another coordinate, so we find $x_1 + x_3 = 1$, $x_1 + x_4 = 1, \ldots, x_1 + x_8 = 1$. Together, these imply that coordinates two through eight are equal: $x_2 = x_3 = \cdots = x_8$. Denoting those equal values by $w$, the vertex at the intersection of these eight facets must have the form

$$v = (1 - w, w, w, w, w, w, w, w, a). \qquad (2.3)$$

Finally, consider that this vertex also lies on the facet defined by $n_1$, and thus is located halfway between the origin and the lattice point $n_1$. This means that

$$x_1^2 + \cdots + x_9^2 = (x_1 - 1/2)^2 + \cdots + (x_8 - 1/2)^2 + (x_9 - a)^2. \qquad (2.4)$$

Again, the quadratic terms cancel, and since $x_9 = a$ we are left with

$$a^2 = -x_1 + 1/4 - \cdots - x_8 + 1/4 = 1 - 6w. \qquad (2.5)$$

This implies that $w = (1 - a^2)/6$; the vertex in Equation (2.3) is listed in Table 1 as type $H_4$.

The remaining vertices listed in the table can be found in the same way, given a set of at least nine facets that intersect at each vertex. To identify all sets of nine facets that intersect at a vertex, we resort to a probabilistic procedure. We select a random vector $c$ uniformly in the unit 9-sphere and solve $\max_{x \in F_9}(c \cdot x)$, where $F_9$ is the Voronoi cell. This is a standard linear program, and the solution is a vertex of $F_9$. Additional vertices are obtained by applying the symmetry group to the found vertex. The process is repeated for multiple random vectors $c$, expanding the vertex set until it saturates, which typically happens after 5–10 000 vectors. Later in the analysis (Section 3) we compute the volume of the convex hull of the vertex set to verify that the set is indeed complete.

The number of vertices in each of the three types of facets can be easily computed from the information given in Table 1. For example the "top" facet of type $F_8^3$, defined by $n_3$, contains all of the vertices with $x_9 = a$, and hence has $16 + 1024 + 1024 + 8960 + 8960 + 7168 + 128 = 27\,280$ vertices. The "vertical" facets of type $F_8^2$ are defined by $n_2$ and have 3484 vertices. The "diagonal" facets of type $F_8^1$ are defined by $n_1$ and have 2454 vertices.

We now shift to a different representation of the facets, identifying each one by the set of vertices that lie in it. These are easily found, by identifying the subset of all vertices whose dot product with the normal vector to the facet is equal to the facet-to-origin distance. From here onward, we also represent lower-dimensional faces in this same way: by the subset of vertices that lie in the face.

To find these lower-dimensional faces, we now carry out a sequence of eight recursive steps, beginning with the $n = 8$-dimensional faces, to obtain the faces in one lower dimension. There are many efficient algorithms for this, for example the general-purpose "diamond-cutting" algorithm of Viterbo and Biglieri[29] or the more recent innovations of Dutour Sikirić and collaborators,[25,30] which can take better advantage of symmetries. However, because our main purpose is to compute $G(a)$, we do not employ these. We simply intersect the two sets of vertices of all pairs of $n$-faces which have common parents, to obtain all $(n-1)$-faces. (If the intersection produces an object with dimension less than $n-1$, it is discarded.) As part of this intersection process in dimension $n$, we construct and save lists of parent faces of dimension $n$ and child faces of dimension $n-1$. Continuing recursively in this way through all the dimensions, we build the complete incidence graph of the Voronoi cell. The number of faces in each dimension is shown in **Table 2**.

**Table 2.** The number of $n$-faces $F_n^t$ of each type $t$, for the $AE_9$ Voronoi cell with $a^2 < 1/2$.

| Dimension $n$ | 0 | 1 | 2 | 3 | 4 | 5 | 6 | 7 | 8 | 9 |
|---|---|---|---|---|---|---|---|---|---|---|
| Types of faces | 1 | 12 | 15 | 13 | 12 | 10 | 7 | 4 | 3 | 1 |
| $F_n^1$ | 93 024 | 218 112 | 584 192 | 645 120 | 430 080 | 358 400 | 57 344 | 10 752 | 256 | 1 |
| $F_n^2$ | | 134 656 | 358 400 | 501 760 | 322 560 | 89 600 | 53 760 | 1344 | 112 | |
| $F_n^3$ | | 107 520 | 197 120 | 369 152 | 286 720 | 80 640 | 8960 | 384 | 2 | |
| $F_n^4$ | | 88 064 | 179 200 | 250 880 | 250 880 | 50 176 | 7168 | 224 | | |
| $F_n^5$ | | 62 720 | 143 360 | 215 040 | 107 520 | 35 840 | 4480 | | | |
| $F_n^6$ | | 53 760 | 125 440 | 150 528 | 98 560 | 17 920 | 2688 | | | |
| $F_n^7$ | | 53 760 | 114 688 | 89 600 | 86 016 | 8960 | 448 | | | |
| $F_n^8$ | | 32 256 | 98 560 | 71 680 | 53 760 | 8960 | | | | |
| $F_n^9$ | | 17 920 | 71 680 | 67 200 | 20 160 | 1120 | | | | |
| $F_n^{10}$ | | 2304 | 40 320 | 50 176 | 16 128 | 896 | | | | |
| $F_n^{11}$ | | 2048 | 35 840 | 50 176 | 10 752 | | | | | |
| $F_n^{12}$ | | 16 | 28 672 | 11 200 | 10 752 | | | | | |
| $F_n^{13}$ | | | 16 384 | 7168 | | | | | | |
| $F_n^{14}$ | | | 1024 | | | | | | | |
| $F_n^{15}$ | | | 1024 | | | | | | | |
| Total faces | 93 024 | 773 136 | 1 995 904 | 2 479 680 | 1 693 888 | 652 512 | 134 848 | 12 704 | 370 | 1 |

## 3. Volume and Second Moment Computations

Once the faces and parent/child (incidence) relationships have been determined, the next step is to compute the volumes, barycenters, and second moments of all $n$-faces. These are computed recursively, beginning in dimension 0 and working upward. See ref. [31] for a useful comparison of different techniques. For our purposes, the most practical method is to construct pyramids from sub-faces, as employed for example in Chapter 21, Theorem 3 of ref. [1]. Unfortunately, we could not employ this directly, because it requires projecting a basepoint (the origin) onto sub-faces in all dimensions. If the vertices are defined by (floating point or rational) numbers, then it is trivial to carry out projections onto sub-faces, by recursively constructing a set of $9-n$ orthonormal vectors to the face $F_n$, for example using the Gram–Schmidt algorithm. But when working analytically with arbitrary $a$, the expressions become increasingly complex. In practice, we were only able to construct projection operators in dimensions six and above.

Since we could not use the projected origin, the natural choice of basepoint was the centroid. A suitable set of recursion formulae is given in Section IV(C) of ref. [29], but the equations there have a number of errors. Similar formulae are given in Section 4 of ref. [25], but are not presented as explicit recursion relations, and we were unsure if the symmetry requirements were satisfied. So here we give the recursion relations that we used, in notation similar to that of ref. [29].

The centroid $C_n^\mu$ of the face $F_n$ is defined by

$$C_n^\mu = \frac{1}{N} \sum_{i=1}^N v_i^\mu, \tag{3.1}$$

where the $v_i^\mu$ are the vertices of the face, $i$ labels the $N$ different vertices, and $\mu = 1, \ldots, 9$ labels the vector components. Because the faces are convex, this point is guaranteed to lie in the interior of the face.

The $n$-volume $V_n$ of a face $F_n$ is defined by

$$V_n = \int_{F_n} d^n x, \tag{3.2}$$

where $d^n x$ denotes the volume element $dx^1 \cdots dx^n$, suitably projected onto the plane of the face. The volume can be computed from the recursion relation

$$V_n = \frac{1}{n} \sum_i h_i V_{n-1}^i, \tag{3.3}$$

where the sub-faces $F_{n-1}^i$ of $F_n$ are labeled by $i$, the volume of the sub-faces is $V_{n-1}^i = V_{n-1}(F_{n-1}^i)$, and $h_i$ is the (positive) height of the centroid of $F_n$ above the plane of the sub-face. The initial condition is $V_0 = 1$.

We use two methods to compute the height $h_i$ to the sub-face $F_{n-1}^i$. In dimensions $n \geq 6$, we first form the projection operator onto the sub-face, which is a matrix $P^{\mu\nu} = \sum \ell^\mu \ell^\nu$. The sum includes $10 - n$ terms; the $\ell^\mu$ are a set of $10 - n$ linearly-independent unit-length mutually-orthogonal vectors normal to the sub-face. The height $h_i$ is then given by the length of $P(C_n - C_{n-1,i})$. In dimensions $n < 6$, we pick $n-1$ vectors $v_j$ that span the sub-face, and compute the Gram determinant, which is the determinant of the $(n-1) \times (n-1)$ square matrix of dot products: $\mathcal{G}_{n-1} = \det[v_j \cdot v_k]$. We then add one more vector $v = C_n - C_{n-1,i}$ to the set and compute the Gram determinant $\mathcal{G}_n$. The height $h_i^2 = \mathcal{G}_n/\mathcal{G}_{n-1}$ is the (squared) volume ratio of the two corresponding parallelotopes.

The barycenter $B_n^\mu$ of a face $F_n$ is defined by

$$B_n^\mu = \frac{1}{V_n} \int_{F_n} x^\mu \, d^n x. \tag{3.4}$$

It is the location which would act as the "center of gravity," or balance point, if the face were constructed from a uniform-density substance, and placed in the $n$-dimensional generalization of a constant gravitational field. We compute the barycenter relative to the centroid of the face, where $O_n^\mu = B_n^\mu - C_n^\mu$ is the offset between them. It is obtained from the recursion relation

$$O_n^\mu = \frac{1}{(n+1)V_n} \sum_i h_i V_{n-1}^i (O_{n-1,i}^\mu + C_{n-1,i}^\mu - C_n^\mu), \tag{3.5}$$

where $h_i$ has the same meaning as before, and we have moved the sub-face label $i$ on $B^\mu$ and $O^\mu$ to the lower position. Note that $O_n^\mu$ vanishes for $n=0$ and $n=1$.

The second moment tensor of a face $F_n$ about the basepoint $x_0$ is defined by

$$U_n^{\mu\nu} = \int_{F_n} (x - x_0)^\mu (x - x_0)^\nu d^n x. \tag{3.6}$$

We always evaluate this about the barycenter of a face, $x_o^\mu = B_n^\mu$; the recursion relation is

$$U_n^{\mu\nu} = \frac{1}{n+2} \times$$
$$\sum_i h_b^i \left[ U_{n-1,i}^{\mu\nu} + \left( B_n^\mu - B_{n-1,i}^\mu \right)\left( B_n^\nu - B_{n-1,i}^\nu \right) V_{n-1}^i \right] \tag{3.7}$$

with $U_0^{\mu\nu} = 0$. Here, in contrast with the previous expressions, $h_b^i > 0$ is the height of the barycenter of $F_n$ above the plane of the $i$th sub-face $F_{n-1}^i$. The trace of this gives the recursion relation for the scalar second moment $U_n = \sum_{\mu=1}^9 U_n^{\mu\mu}$ about the barycenter. In many instances, as described in ref. [25], these recursive sums can be simplified, so that there is one term for each congruence class of sub-faces, rather than one term per sub-face.

The Supporting Information includes face catalogs for the different ranges of $a$. The catalogs contain complete lists of formulae for volume, barycentric and centroid heights, and second moments, of all face types in all dimensions. From these, the formulae we give here can be reconstructed. For example, the three types of eight-faces have volume

$$V(F_8^1) = \sqrt{a^2 + 2} \left( \frac{a^{15}}{64} - \frac{a^{13}}{30} + \frac{7a^9}{180} - \frac{7a^5}{180} + \frac{a}{30} \right),$$

$$V(F_8^2) = \sqrt{2} \left( \frac{-a^{15}}{28} + \frac{8a^{13}}{105} - \frac{4a^9}{45} + \frac{4a^5}{45} + \frac{a}{15} \right), \text{ and}$$

$$V(F_8^3) = -a^{16} + \frac{32a^{14}}{15} - \frac{112a^{10}}{45} + \frac{112a^6}{45} - \frac{32a^2}{15} + 1. \tag{3.8}$$

The heights of these faces from the centroid of the Voronoi cell about the origin $F_9^1$ are easily calculated, and are (respectively)

$$h_1 = \sqrt{a^2 + 2}/2, \quad h_2 = \sqrt{2}/2, \text{ and } h_3 = a, \tag{3.9}$$

from which one obtains $V(F_9^1) = 2a$. This also follows immediately from the determinant of the generator matrix of Equation (1.2), providing a simple consistency check and confirming that all vertices had been found.

The symbolic calculations which lead to these results are too time-consuming to repeat for every face. To speed the process up, we exploit symmetry: the volume calculations are done only once for each face type in dimension $n = 1, \ldots, 8$, and then used for all similar faces in that dimension. One way to identify equivalent faces would have been to search for the required symmetry transformation, but this is challenging to program and implement. So we used two other methods. First, we computed the volumes numerically for $a = 0.573$, where the values are distinct, apart from dimension $n = 3$, where there are two different faces that have identical three-volumes:

$$V(F_3^2) = V(F_3^4) = \frac{a\sqrt{12a^2 + 7}(3 - 2a^4)}{72}. \tag{3.10}$$

Fortunately these are trivially distinguished, because the first has six sub-faces, and the second has seven sub-faces. We also checked that we could uniquely identify face types by making a hash from the hierarchy of child counts.

As shown in Table 2, the $AE_9$ Voronoi cell contains 78 distinct types of faces. For all but one type, the sub-faces of a given type are all at the same barycentric height and separation. The exception is $F_2^5$, whose two sub-faces of type $F_1^6$ have different values of barycenter separation $|B_2 - B_1|$. Both are listed in the face catalog.

## 4. Phase Transitions for $a^2 > 1/2$

While the optimal lattice quantizer of the form (1.2) has $a^2 < 1/2$, we have also studied how the lattice behaves for larger values. As $a$ increases from zero, the Voronoi cell evolves through different "phases."

Within each phase, corresponding to the L-type domains of ref. [5], the cell has a constant number of facets, vertices, and face equivalence classes, and the combinatorial structure (by which we mean the network topology of the face incidence graph) is invariant. This is shown in **Table 3**. (Note that the structure of the $a = 1$ case is explicitly calculated in Section 7 of ref. [32], setting $n = 9$). Varying $a$ smoothly across a phase transition, the vertices merge together or split apart, and the "velocity" $dx^\alpha/da$ of some of the vertices changes direction.

Previously, we studied phase A of Table 3, because it is relevant for the conjectured optimal lattice quantizer in 9 dimensions, which has $a \approx 0.573$. Consider instead, for example, phase B, for which $1/2 < a^2 < 1$. Although the number of vertex equivalence classes is the same as for $a^2 < 1/2$, four of the classes

**Table 3.** Voronoi cell structure for increasing $a$.

| Phase | | Vertices | Classes | Facets | Classes |
|---|---|---|---|---|---|
| A: | $0 < a^2 < 1/2$ | 93 024 | 16 | 370 | 3 |
| | $a^2 = 1/2$ | 54 496 | 14 | 370 | 3 |
| B: | $1/2 < a^2 < 1$ | 66 144 | 16 | 370 | 3 |
| | $a^2 = 1$ | 8160 | 7 | 370 | 3 |
| C: | $1 < a^2 < 2$ | 9344 | 9 | 370 | 3 |
| | $a^2 = 2$ | 7138 | 7 | 368 | 2 |
| D: | $2 < a^2$ | 7266 | 7 | 368 | 2 |

$H_9$, $H_{10}$, $H_{12}$, and $H_{13}$ are modified from those given in Table 1. Their respective coordinates take the following new forms:

$$\left(-w, w, w, \frac{1}{2}, \frac{1}{2}, \frac{1}{2}, \frac{1}{2}, \frac{1}{2}, 0\right), \quad w = a^2 - \frac{1}{2},$$
$$(1 - w, w, w, 0, 0, 0, 0, 0, a), \quad w = 1 - a^2,$$
$$\left(0, 0, 0, 0, 0, \frac{1}{2}, \frac{1}{2}, \frac{1}{2}, w\right), \quad w = \frac{a}{2} + \frac{1}{4a}, \text{ and}$$
$$\left(0, 0, 0, \frac{1}{2}, \frac{1}{2}, \frac{1}{2}, \frac{1}{2}, \frac{1}{2}, w\right), \quad w = \frac{a}{2} - \frac{1}{4a}. \tag{4.1}$$

Phases A and B have the same numbers of faces (see Table 2) in dimensions 6 through 9, but different numbers of faces in dimensions 5 and below (see face catalogs in the Supporting Information).

We calculate the volumes and second moments for all four phases, in the same way as previously described; full details are provided in the accompanying face catalogs (see Supporting Information). For example in phase B, for $1/2 < a^2 < 1$, the unnormalized second moment is

$$U_B =$$
$$\frac{121}{12\,150}a^{19} - \frac{92}{1\,215}a^{17} + \frac{32}{135}a^{15} - \frac{152}{405}a^{13} + \frac{112}{405}a^{11}$$
$$- \frac{28}{675}a^9 + \frac{28}{405}a^7 - \frac{152}{405}a^5 + \frac{181}{270}a^3 + \frac{1\,393}{1\,215}a + \frac{1}{48\,600}a^{-1}. \tag{4.2}$$

Since their domains in $a$ are disjoint, we cannot immediately compare $U_B$ to the corresponding equation for $U_A$ given in Equation (1.4) for $a^2 < 1/2$. However, both functions may be trivially extended to the entire real $a$ axis by analytic continuation, and then directly compared.

The result is a beautiful sequence of simple relationships. The volume $V$ is the determinant of the generator matrix Equation (1.2), so the same volume formula $V = 2a$ holds for all four phases, and all of its derivatives are continuous. But the second moments change form at the transition points, with differences:

$$U_B - U_A = \frac{(2a^2 - 1)^{10}}{97\,200a^2} V,$$
$$U_C - U_B = \frac{(1-a^2)^9 (2a^2 + 3)}{405a^2} V, \text{ and}$$

$$U_D - U_C = -\frac{(a^2 - 2)^{10}}{24\,300a^2} V. \tag{4.3}$$

It would be unfair to call these "jumps," because the transitions are extremely smooth. A physicist would describe these as ninth or tenth order phase transitions, because at the transition points, the Taylor series of $U/V$ agree to this order. There is probably a simple geometrical explanation for this behavior.

The second moment tensors $U^{\mu\nu} = \alpha \delta^{\mu\nu} + \beta z^\mu z^\nu$ have similar behavior across the transitions:

$$\beta_B - \beta_A = \frac{(2a^2 - 1)^9 (16a^2 + 1)}{77\,760a^2} V, \tag{4.4}$$

$$\beta_C - \beta_B = -\frac{(1 - a^2)^8 (32a^4 + 43a^2 + 6)}{648a^2} V, \text{ and} \tag{4.5}$$

$$\beta_D - \beta_C = -\frac{(a^2 - 2)^9 (4a^2 + 1)}{9720a^2} V. \tag{4.6}$$

The corresponding formulae for $\alpha$ may be easily obtained from the trace constraint $U = 9\alpha + \beta$.

## 5. Discussion

More than three decades have passed since the publication of Conway and Sloane's wonderful book on sphere packings, lattices and groups,[1] and more than two decades since the third edition appeared. It is sobering to compare the plot they provide (see Figure 2.9 of Chapter 2 in ref. [1]) of the best quantizers known in dimensions $n \leq 24$ with the current state of affairs. Since that plot was published 37 years ago,[36] there have been only a few additions, which are shown in **Figure 2**. The main progress has been in dimensions 8–12 and can be seen more clearly in the enlargement **Figure 3**.

In dimension 9, there has been the numerical discovery of the $AE_9$ lattice by Agrell and Eriksson,[19] who also identified the (non-lattice) tessellation $D_9^+$ as a superior quantizer. Dutour Sikirić[36] subsequently calculated the moments of $D_9^+$ analytically, using the methods of [25], obtaining $G(D_9^+) = 924\,756\,607/13\,005\,619\,200 = 0.0711043\ldots$, which is in good agreement with the numerical results reported in ref. [19]. Table 5 of ref. [25] also reports exact values for the laminated lattice $\Lambda_9$ and its dual, $G(\Lambda_9) = 151\,301/2\,099\,520 = 0.0720645\ldots$ and $G(\Lambda_9^*) = 1\,371\,514\,291/19\,110\,297\,600 = 0.0717683\ldots$, as well as exact values for the Coxeter lattices $G(A_9^2) = 5^{-1/9} 2^{-8/9} 2\,120\,743/13\,271\,040 = 0.0721668\ldots$, and $G(A_9^5) = 2^{-1/9} 5^{-8/9} 8\,651\,427\,563/26\,578\,125\,000 = 0.0720790\ldots$.

In dimension 10, the exact value for $G(D_{10}^+) = 4\,568\,341/64\,512\,000 = 0.0708138\ldots$ was computed in ref. [25]. This is also in good agreement with the value found numerically in ref. [19], which numerically converged on this lattice as the optimal lattice quantizer in ten dimensions.

In dimension 11, ref. [25] has provided exact values for the quantizer constants of $A_{11}^2$, $A_{11}^4$, and $A_{11}^3$; the last of these is the current record-holder (see Figure 3).

In dimension 12, $K_{12}$ remains the best lattice quantizer, although there has been some progress: ref. [25] calculates an exact value for $G(K_{12})$, which was previously only estimated by Monte Carlo integration, and also calculates $G(D_{12}^+)$, which is larger.

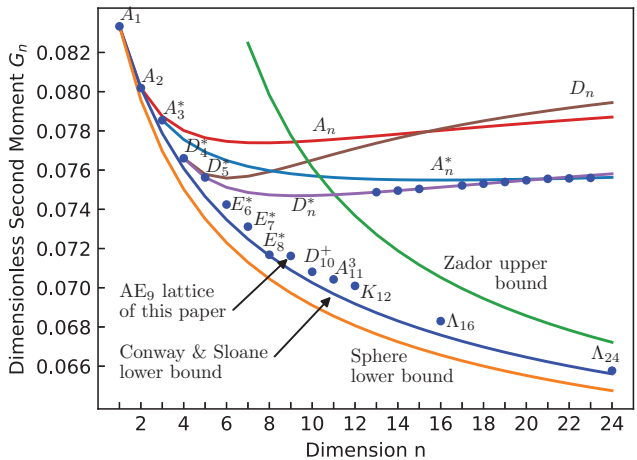

**Figure 2.** The best lattice quantizers currently known are shown as dots, in dimensions $n \leq 24$. Also shown are the conjectured lower bound of Conway and Sloane (see Equation (4) in ref. [33]), the lower-bound from the interior of a sphere (see Equation (2) in ref. [33]), and Zador's upper bound.[16,34] In dimensions $13 \leq n \leq 15$ and $17 \leq n \leq 23$, the best quantizers currently known are $D_n^*$ and $A_n^*$, which are above Zador's upper bound.

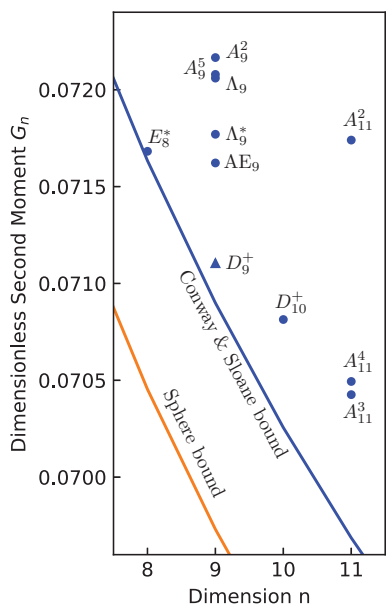

**Figure 3.** An enlargement of Figure 2 around dimension 9, showing other good lattice quantizers. In dimension 9, the second-best lattice quantizer, $\Lambda_9^*$, is a special case of the lattice studied in this paper, corresponding to $a = 1/2$. The plot includes one non-lattice quantizer, the tessellation $D_9^+$ discussed in detail in ref. [19].

In dimensions $13 \leq n \leq 24$, with the exceptions of $n = 16$ and $n = 24$, the best lattice quantizers that we currently know lie above the upper bound given by Zador.[16,34] These are $D_n^*$ for $13 \leq n \leq 20$ and $A_n^*$ for $21 \leq n < 24$. Zador's bound establishes that in these dimensions there must be superior quantizers, but since the proof is not constructive, their identity is unknown; it is not even known if the optimal quantizers are lattices.

## 6. Further Applications

The $AE_9$ lattice has some practical interest. For example, a search for pulsars in eccentric binary orbits has nine non-trivial parameter-space dimensions (see Sections 4.1 and 6 in ref. [37]). Using the optimal quantizer as a parameter-space grid would minimize the number of "lost" detections.[15] But the more surprising aspect of $AE_9$ is how it suddenly appears in dimension 9, after a long sequence of well-known classical lattices. Is this a hint of what happens in (some) higher dimensions?

Happily, our methods can be used to calculate $G$ for other one-parameter families of lattices, and to search for optimal quantizers in higher dimensions. They are particularly suitable for building laminated lattices, which are constructed by "stacking copies" of an $(n-1)$-dimensional lattice in the $n$th dimension.

To make an $n$-dimensional laminated lattice (see Chapters 5 and 6 in ref. [1] and refs. [30,38]), we start with a generator $B_{n-1}$ for the lower-dimensional lattice, and an $(n-1)$-dimensional vector $r$, which is the component of the stacking offset in the $(n-1)$-plane. This is often taken to point toward a "deep hole" in the $(n-1)$-lattice (the vertex of the Voronoi cell most distant from the origin). We then construct an $n$-dimensional generator matrix

$$B = \begin{bmatrix} B_{n-1} & 0 \\ r & a \end{bmatrix}, \quad (6.1)$$

where $a$ is a positive real number, which has zeros above it. The $AE_9$ generator Equation (1.2) has exactly this form because $B_{n-1}$ is the generator for $D_8$ and $r$ points to a deep hole. The value of $a$ determines the distance between the shifted lattice copies, in the direction orthogonal to their plane.

With this assumed form, it should be possible to calculate $G$ as we have done here. In Table 1, the coordinates of all vertices of the Voronoi cell are quadratic functions of $a$. The situation is similar for the laminated construction of Equation (6.1), where the parameter $a$ sits in the bottom right corner, with vanishing entries above it. This form of $B$ implies that all but the final coordinate of the Voronoi cell vertices take the quadratic form $c + \bar{c}a^2$; the final coordinate $x_n$ has the form $x_n = c/a + \bar{c}a$. Here $c$ and $\bar{c}$ are constants, of which at least one is non-zero. For example, for the $AE_9$ lattice, compare Table 1 and Equation (4.1): the final coordinate has $c = 0$ for $a^2 < 1/2$, and $c \neq 0$ for $1/2 < a^2$.

This is enough to infer the functional form of $V = V_n(a)$ and $U^{\mu\nu} = U_n^{\mu\nu}(a)$. The volume $V$ is given by the determinant of $B$, and is therefore proportional to $a$. The second moment $U^{\mu\nu}$ may be found by decomposing the Voronoi cell into $n$-simplices. There are many ways to do this, as elaborated in ref. [31]. For now, suppose that this is done recursively: each face in dimension $n$ is written as a union of $n$-simplices, formed by adding the centroid of the face as a new vertex to each of the $(n-1)$-simplices used to decompose the lower-dimensional faces. In this way, the Voronoi cell is written as a sum of $n$-simplices, each of which has $n+1$ vertices. One of those vertices is the origin, $n-2$ of those vertices are centroids of lower-dimensional faces, and the final two are drawn from the vertices of the Voronoi cell.

Such a decomposition leads to a combinatorial explosion because the number of simplices grows very rapidly with dimension. For example, if the $AE_9$ Voronoi cell with $a^2 < 1/2$ is decomposed in this way, it is expressed as the union of 45 344 194 560 nine-simplices. But to determine the functional form of $U(a)$, it is only necessary to imagine that this has been done, but not to carry it out in practice.

Let the $n + 1$ vertices of one of the simplices be the vectors $0^\mu, \alpha^\mu, \ldots, \beta^\mu$. Then the volume of the simplex is the determinant of the $n \times n$ square matrix

$$V_n = \frac{1}{n!} \det \begin{bmatrix} \alpha^1 & \cdots & \alpha^n \\ \vdots & \ddots & \vdots \\ \beta^1 & \cdots & \beta^n \end{bmatrix}, \qquad (6.2)$$

where the rows are ordered to give a positive volume, and the superscripts denote the $n$ components of the vectors. Two of the rows are vertices of the Voronoi cell, and the remaining rows are centroids of the sub-faces in dimensions 2 to $n-1$, defined in Equation (3.1). Since the centroids are averages of the Voronoi cell vertices, they have the same functional form as the vertices. Hence, all but the final column of Equation (6.2) have the form $c + \bar{c}a^2$; the final column has the same form, but with an additional factor of $1/a$. (For $AE_9$ with $a^2 < 1/2$ only the term linear in $a$ appears because $c = 0$.) Consequently, the volume of each simplex takes the form $V_n = P(a^2)/a$, where $P$ is polynomial of maximum order $n$ in $a^2$.

The second moment tensor $U^{\mu\nu}$ about the origin for an $n$-simplex is easily calculated. After normalization by the volume, one has

$$I^{\mu\nu} = U^{\mu\nu}/V = \frac{(\alpha^\mu + \cdots + \beta^\mu)(\alpha^\nu + \cdots + \beta^\nu) + \alpha^\mu \alpha^\nu + \cdots + \beta^\mu \beta^\nu}{(n+1)(n+2)}, \qquad (6.3)$$

where we have assumed that one of the simplex's vertices is at the origin, as is the case here. (The trace of this equation reduces to the scalar second moment given in Chapter 21, Theorem 2 of ref. [1].) By a reasoning similar to that used in the previous paragraphs, $I^{\mu\nu}$ has components of the form $a^{-2} S(a^2)$ where $S$ are polynomials of maximum order 3. Hence, adding the un-normalized second moment tensors $U^{\mu\nu} = V I^{\mu\nu}$ of all of the simplices, we find that the total un-normalized second moment tensor of the Voronoi cell has components of the form $P(a^2)/a^3$, where $P$ is a polynomial of maximum order $n + 3$.

Making use of these forms, the dimensionless (scalar) second moment can be written as

$$G(a^2) = \frac{1}{n} \frac{U}{V^{1+2/n}} = \frac{1}{n} P(a^2)(a^2)^{-(2n+1)/n}. \qquad (6.4)$$

The extrema of $G$ satisfy $G'(a^2) = 0$, where $' = d/da^2$, which implies

$$na^2 P'(a^2) - (2n+1)P(a^2) = 0. \qquad (6.5)$$

This is a polynomial equation in $a^2$ of maximum order $n + 3$, which means that one may use this technique to perturb any lattice whose Voronoi cell can be calculated, and to identify the extrema. If the matrix $B_{n-1}$ and vector $r$ are rational, this also demonstrates that the extremal values of $a$ are algebraic numbers.

Existing techniques can compute the quantizer constants for specific lattices using rational arithmetic.[25,29,30] To find the polynomial $P$, it is sufficient to compute $U$ or $G$ for $n + 3$ distinct rational values of $a$. Since $P$ is polynomial, one can use these values to then obtain a set of linear equations for the coefficients. Solving these yields the polynomial for $U$, and hence the one-parameter family of solutions.[39]

The program we are describing, to generate one-parameter families of lattices, Voronoi cells, and quantizer constants, may provide a useful technique for generating better quantizers in higher dimensions, and understanding their structure. If needed, the choice of starting generators and the direction to perturb them can be guided by further numerical studies like those of ref. [19].

## Supporting Information

Supporting Information is available from the Wiley Online Library or from the author.

## Acknowledgements

The authors are grateful for help from the SymPy developers, in particular Oscar Benjamin and Aaron Meurer, in understanding how to automate simplification of expressions involving absolute values, such as $\sqrt{(2a^2 - 1)^2} = |2a^2 - 1| = 1 - 2a^2$. They also thank Mathieu Dutour Sikirić for confirming that his face counts for $\Lambda_9^*$ match those that we found for $AE_9$,[26] and Daniel Pook-Kolb for several helpful comments and corrections.

Open access funding enabled and organized by Projekt DEAL.

## Conflict of Interest

The authors declare no conflict of interest.

## Data Availability Statement

The data that supports the findings of this study are available in the Supporting Information of this article.